\documentclass{elsart}
\usepackage{graphicx}
\usepackage{epsfig}

\begin{document}
\begin{frontmatter}
\title{Coulomb corrections for quasielastic (e,e') scattering:
eikonal approximation}
\author{Andreas Aste\corauthref{cor}}
\corauth[cor]{Corresponding author.}
\ead{aste@quasar.physik.unibas.ch},
\author{Kai Hencken, J\"urg Jourdan, Ingo Sick, Dirk Trautmann}
\address{Departement f\"ur Physik und Astronomie,
Universit\"at Basel, 4056 Basel, Switzerland}
\date{August 5, 2004}
\begin{abstract}
We address the problem of including Coulomb distortion
effects in inclusive qua\-si\-elas\-tic $(e,e')$ reactions
using the eikonal approximation.
Our results indicate that Cou\-lomb corrections may become
large for heavy nuclei for certain kinematical regions.
The issues of our model are presented in detail and the results are
compared to calculations of the Ohio group, where
Dirac wave functions were used both for electrons and
nucleons. Our results
are in good agreement with those obtained by exact calculations.
\end{abstract}

\end{frontmatter}

\section{Introduction} 
Nucleon knockout by electron scattering provides a powerful probe of the  dynamics of nucleons in the nuclear medium.
The transparency of the nucleus with respect to
the electromagnetic probe makes it
possible to study the entire nuclear volume.
For light nuclei, the weakness of the electromagnetic
interaction allows for a separation of the soft Coulomb distortion
of the electron scattering process from the hard scattering event
in which, to a very good approximation,
a single virtual photon transfers energy and momentum to the nuclear
constituents. Since the kinematic conditions
of electron scattering can be varied easily,
different aspects of the reaction mechanism can be tested.
Under conditions in which a single nucleon receives
most of the energy and momentum transfer, the quasifree electron-nucleon
scattering process is emphasized.

Nucleon knockout from heavier nuclei is used to measure the
single-nucleon spectral function in complex nuclei.
At low excitation energy, exclusive measurements to discrete states of the
residual nucleus provide information on quasiparticle properties,
such as binding energies,
spectroscopic factors, spreading widths and momentum distributions.
The specific models used to extract the structure information
can be tested by comparing parallel with
nonparallel kinematics or by varying the ejectile energy.
The spatial localization of specific orbitals also provides
some sensitivity to possible density-dependent modifications
of the electromagnetic properties of bound nucleons.
The azimuthal dependence
of the cross section and the recoil polarization of the
ejectile can provide detailed tests of the reaction mechanism,
which may be useful in delineating the role of two-body currents
or in testing off-shell models of the current operator.
Measurements at larger missing energy can provide information
on the deep-hole spectral function or on multinucleon currents.
Measurements at large missing momentum
are sensitive to short-range and tensor correlations
between nucleon pairs.

For heavier nuclei Coulomb corrections (CC) may become very large
and affect the measured cross sections; this needs to be accounted for,
if one aims at a quantitative interpretation of data.
Here, we concentrate on the {\em{inclusive}} quasielastic
scattering process $(e,e')$, where only the scattered electron is
observed. We will model this process as a
knockout reaction where the nucleons are hit
by the virtual photon emitted by the scattered electron.

Inclusive scattering provides information on a number
of interesting nuclear properties:
\begin{description}
\item{-}The width of the quasielastic peak allows a
{\em{dynamical}} measurement of the nuclear Fermi momentum
\cite{Whitney74}.
\item{-}The tail of the quasielastic peak at low energy loss
and large momentum transfer gives information on high-momentum
components in nuclear wave functions \cite{Benhar94a}.
\item{-}The integral strength of quasielastic scattering,
when compared to sum rules, tells us about the reaction mechanism
and eventual modifications of nucleon form factors in the nuclear medium
\cite{Jourdan96a}.
\item{-}The scaling properties of the quasielastic response allows
to study the reaction mechanism \cite{Day90}.
\item{-}Extrapolation of the quasielastic response to $A=\infty$
provides us with a very valuable observable for infinite nuclear matter
\cite{Day89}.
\end{description}
For heavier nuclei, these questions obviously can
only be addressed once the Coulomb distortion of the
electron waves is properly dealt with.

In this paper, we present an approximate treatment of electron
CC for inclusive quasielastic $(e,e')$ reactions
modeled as a nucleon knockout process.
In the plane-wave Born approximation (PWBA), the
electrons are described as plane Dirac waves, which is a
poor approximation for heavy nuclei with strong
Coulomb fields. In a better approach,
called eikonal distorted wave Born approximation
(eDWBA), we use electron waves which are
distorted by an additional phase and a change in
the amplitude.
This phase shift and the modification of the amplitude account
for the enhanced momentum and a focusing effect which occurs
when the electron approaches the strongly
attractive nucleus.

Calculations with exact Dirac wave functions
have been performed by Kim {\emph{et al.}} \cite{Jin1} in the Ohio group
and Udias {\emph{et al.}} \cite{Udias,Udias2};
the present eikonal
approximation has the advantage that it is relatively simple to implement
and that it avoids large computational costs.
The eikonal method and its higher approximations, which can
be obtained from an iterative procedure, is expected to lead to
an asymptotic, rather than a convergent expansion \cite{Yennie64}.
Therefore, its use for the calculation of exclusive cross sections
may be problematic according to Giusti and Pacati
\cite{Giusti,Giusti2}, but the good agreement for the
inclusive cross section with the exact
calculations by Kim {\emph{et al.}} seems to justify the
method in this case. Additionally, the eikonal approximation
used in \cite{Giusti,Giusti2} is not equivalent
to the approach used in this paper, since we do not calculate
the electron wave phase shift from
an expansion around the center of the nucleus.
For exclusive cross sections, not only
exact wave functions for the electrons are needed, but also
an accurate description of the proton or neutron wave
functions. In the inclusive case the fine details of the
nuclear structure and final state interaction
are not important, and the description of the unobserved
knocked-out hadron can be rather cursory.

Various approximate treatments have been proposed in the past
for the treatment of CC
\cite{Giusti,Yennie54,Lenz,Knoll,Traini88,Traini95,Kosik,Rosenfelder,Rosenfelder80},
and there is an extensive literature
on the eikonal approximation
\cite{Sucher,Blankenbecler,Wallace1,Wallace2,Abarbanel}.
In particular it has been shown that at lowest order, an expansion
of the electron wave function in $\alpha Z$ leads to the well known
effective momentum approximation (EMA) \cite{Yennie64}, which is
explained in Sect. 2.2.
The EMA has been used to correct Coulomb distortions
for elastic electron scattering data or inelastic data to low lying
states, although exact descriptions are readily available.
For quasielastic $(e,e')$ scattering a comparison
of EMA calculations with numerical results from the 'exact' DWBA
calculation \cite{Onley92,Onley94} indicates a failure of the EMA
\cite{JourdanWorkshop}. 

We point out that the model of the nuclear structure used
in this paper is relatively simple, since our focus is mainly on
the electronic part of the problem, i.e. the influence of the
Coulomb field on the electron wave function. 
Still, our simple model leads to satisfactory values for inclusive
cross sections which are sufficient for our purpose.
We also find that the EMA fails to account
for CC for quasielastic $(e,e')$.
This observation is discussed in detail in Sect. 3.

\section{Models and approximations}
\subsection{Effective momentum approximation}
The accurate description of the electrostatic potential of
the nucleus is an important ingredient for the calculation
of CC. In a simplified classical picture,
the electron hits mostly the outer region of the
nucleus. This implies that for heavy nuclei like lead with a radius of
approximately $6 \, \mbox{fm}$,
the kinetic electron energy is increased
in this region by about 20 MeV, which is a non negligible modification
when the electron energy lies in the range of a few hundreds
of MeV.
In the most naive picture, the effective momentum approximation,
one typically assumes that the nucleus is a
homogeneously charged sphere with equivalent radius $R_e$.
For highly relativistic particles in a potential $V$ with
asymptotic momentum $\vec{k}$,
one may neglect the mass of the
particle ($|E-V|, |\vec{k} \, | \gg m$),
such that the energy-momentum relation reduces to
($\hbar=c=1$)
\begin{equation}
(E-V)^2=\vec{k}^{\, 2} + m^2 \, \rightarrow \, E-V=k, \, k=|\vec{k}|.
\end{equation}
Then the momentum shift $\Delta k$ of a highly relativistic electron in
the region where it interacts with the nucleus
follows from the potential energy of the electron
inside the nucleus
($C=1$ at the surface or $C=3/2$ in the center)
\begin{equation}
\Delta k = C \frac{\alpha Z}{R_e}, \quad C=1 \ldots 3/2, \quad
k^{\mbox{\scriptsize{eff}}}_{i,f}=k_{i,f}+
\Delta k \, . \label{deltaeff}
\end{equation}
The standard EMA uses effective momenta corresponding to the
central value of the Coulomb potential (C=3/2).
The EMA cross section of the considered process
is then calculated by using the plane wave approach,
but with the electron momenta replaced by their corresponding
effective values. In this way one accounts for the fact
that the electron wave length is reduced in the relevant nuclear region.
But due to the attractive Coulomb potential, the modulus of the
initial (final) electron wave is also enhanced by a factor
of $\sim F_i$ ($F_f$) inside the nuclear volume.
The cross section is therefore multiplied additionally by a factor
$F_i^2=(k^{\mbox{\scriptsize{eff}}}_{i}/k_i)^2$
which accounts for the focusing of the incoming electron wave
in the nuclear center. 
The cross section is not multiplied by $F_f^2= (k^{\mbox{\scriptsize{eff}}}_{f}/k_f)^2$,
because this factor is already contained in the artificially
enhanced phase space factor of the outgoing electron.

In order to be more explicit, we mention that
in the plane wave Born approximation, the cross section for
inclusive quasielastic electron scattering can be written by the help of
the total response function $S_{total}$ as
\begin{equation}
\frac{d^2 \sigma_{_{PWBA}}}{ d \Omega_f d\epsilon_{f}}=
\sigma_{Mott} \times S_{total}(|\vec{q} \, |,\omega,\Theta_e),
\label{Mott1}
\end{equation}
where the Mott cross section is given by
\begin{equation}
\sigma_{Mott}=4 \alpha^2 \cos^2(\Theta_e/2) \epsilon_f^2/q_\mu^4.
\label{Mott2}
\end{equation}
Here, $\alpha$ is the fine-structure constant,
$\Theta_e$ is the electron scattering angle,
$\vec{q}=\vec{k}_i-\vec{k}_f$ is the momentum transfer given
by the initial and final electron momentum,
$\omega=\epsilon_i-\epsilon_f$ is the energy transfer given
by the initial and final electron energy,
and the four-momentum transfer squared is given by
$q_\mu^2=\omega^2-\vec{q}^{\, 2}$.

The Mott cross section remains unchanged
when it gets multiplied by the EMA focusing factors
and the momentum transfer $q_\mu^4$ is replaced by its
corresponding effective value, i.e.
$F_i^2 F_f^2/q_{\mu,\mbox{\scriptsize{eff}}}^4 =
1/q_\mu^4$. Therefore, the EMA cross section can also be obtained
from (\ref{Mott1}) by leaving the Mott cross section
unchanged and by replacing $S_{total}(|\vec{q} \, |,\omega,\Theta_e)$
by the effective value
$S_{total}(|\vec{k}_i^{\mbox{\scriptsize{eff}}}-
\vec{k}_f^{\mbox{\scriptsize{eff}}}|, \omega,\Theta_e)$.

\subsection{Eikonal approximation}
We can take the local change in the momentum of the incoming particle
with momentum $\vec{k}_i=k_i \hat{k}_i$ into account
approximately by modifying the plane wave describing the
initial state of the particle by the eikonal phase $\chi_1(\vec{r})$
\begin{equation}
e^{i\vec{k}_i\vec{r}} \, \rightarrow \, e^{i \vec{k}_i\vec{r}+i\chi_1(\vec{r})} \, ,
\end{equation}
where
\begin{equation}
\chi_1(\vec{r})=-\int \limits_{-\infty}^{0} V(\vec{r}+
\hat{k}_i s) ds=
-\int \limits_{-\infty}^{z} V(x,y,z') dz' \,
\end{equation}
if we set $\vec{k}_i=k_z^i \hat{{\bf{e}}}_z$.
As desired, the $z$-component of the momentum then becomes
\begin{equation}
p_z=-i \partial_z e^{i k_z^i z+i\chi_1}=
(k_z^i-V)e^{i k_z^i z+i\chi_1}.
\end{equation}
The final state wave function is constructed analogously
\begin{equation}
e^{i \vec{k}_f\vec{r}-i\chi_2(\vec{r})},
\end{equation}
where
\begin{equation}
\chi_2(\vec{r})=-\int \limits_{0}^{\infty} V(\vec{r}+
\hat{k}_f s') ds' \, .
\end{equation}
For the sake of simplicity, we consider spinless electrons
in Sects. 2.1-2.4.
In our actual calculations spin is included.
The spatial part of the free electron current
(which interacts via photon exchange with the
particles inside the nucleus)
\begin{equation}
\vec{j}_{_{PW}}=-ie[e^{-i \vec{k}_f\vec{r}} \vec{\nabla}
e^{i \vec{k}_i\vec{r}}-(\vec{\nabla}e^{-i \vec{k}_f\vec{r}})
e^{i \vec{k}_i\vec{r}}]= e(\vec{k}_i+\vec{k}_f)e^{i(\vec{k}_i-
\vec{k}_f)\vec{r}} 
\end{equation}
is replaced by
\begin{displaymath}
\vec{j}_{_{EIK}}=-ie[e^{-i \vec{k}_f\vec{r}+i\chi_2(\vec{r})} \vec{\nabla}
e^{i \vec{k}_i\vec{r}+i\chi_1(\vec{r})}-
(\vec{\nabla}e^{-i \vec{k}_f\vec{r}
+i\chi_2(\vec{r})}) e^{i \vec{k}_i\vec{r}+i\chi_1(\vec{r})}]=
\end{displaymath}
\begin{equation}
e(\vec{k}_i+\vec{k}_f+\vec{\nabla} \chi_1-\vec{\nabla} \chi_2)
e^{i(\vec{k}_i-\vec{k}_f)\vec{r}+i(\chi_1+\chi_2)} \, ,
\end{equation}
where $e$ is the charge of the electron.
The spatial part of the electron current now contains the
additional eikonal phase, and the prefactor 
contains gradient
terms of the eikonal phase which represent essentially
the change of the electron momentum due to the attraction of
the electron by the nucleus.

So far we have only considered the modification of the
phase of the wave function, and for many applications this is a sufficient
approximation. It has been applied to elastic high energy
scattering of Dirac particles in an early paper by Baker \cite{Baker}.
However, the method leads, e.g. for quasielastic
scattering of electrons on lead with
initial energy $\epsilon_i=300$ MeV and energy transfer $\omega=100$ MeV,
to errors up to 50\% in the cross sections.
The reason is that also the amplitude of
the incoming and outgoing particle
wave functions is changed due to the Coulomb attraction,
as mentioned above.
This fact can be related to the classical observation
that an ensemble of negatively charged test particles approaching
a nucleus is focused due to its attractive potential.
Below, we give a simple classical treatment of this fact
which illustrates the basic properties of the phenomenon.
A semiclassical discussion of the focusing factor can also be found
in \cite{Yennie64}.

\subsubsection{The focusing factor} 

The focusing factor can be derived approximately from a classical
toy model according to Fig. ({\ref{figfoc}).
We consider the trajectories of
an ensemble of highly relativistic particles approaching the
nucleus in the center of the coordinate system
in $z$-direction with an impact parameter between $b_0$
and $b_0+db_0$. The longitudinal velocity of the particles can be taken
as the speed of light, since the particles are
highly relativistic and the change of the velocity in transverse
direction causes only a second order effect to the longitudinal component.
Therefore we have $z(t)=t$, $r(t)=\sqrt{b_0^2+t^2}$, keeping in mind
that the impact parameter can be considered as constant at this stage
of our approximation.
The density of the particles at $z$ is increased by a
focusing factor $f$
which is given by the ratio of the area of two annuli with radii
$b_0, b_0+db_0$ and $b(b_0,z), b(b_0+db_0,z)$:
\begin{equation}
f^{-1}(b_0,z)=\frac{\partial b(b_0,z)}{\partial b_0}
\frac{b(b_0,z)}{b_0}.
\end{equation}
We will calculate now $b(b_0,z)$ for the screened potential
\begin{equation}
V_{s}(r)=-\frac{\alpha Z}{\sqrt{r^2+R^2}} ,
\end{equation}
since in this case a simple study of the problem and
its most important properties is possible.

\begin{figure}
\begin{center}
        \includegraphics[width=7.8cm]{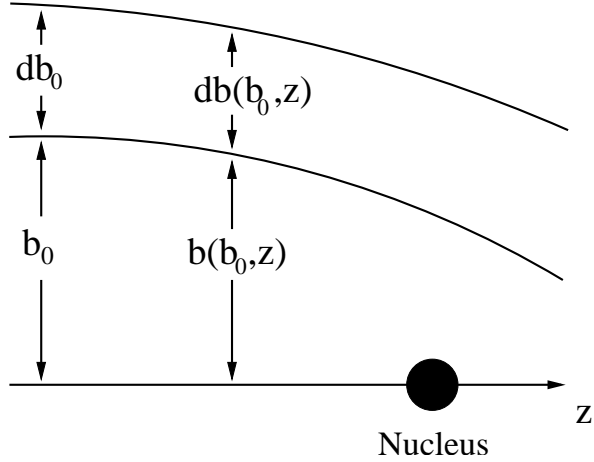}
        \caption{Electrons incident on an
         nucleus with impact parameter $b_0$.}
        \label{figfoc}
\end{center}
\end{figure}

The force $F_T$ acting on the particle in transverse direction is
given by
\begin{equation}
F_T = -\frac{b}{r} \frac{\partial V(r)}{\partial r} \sim
-\frac{\alpha Z b_0}{\sqrt{r^2+R^2}^3},
\end{equation}
whereas the
"transverse mass" is given by the energy $E$ $(c=1)$.
Therefore we obtain for the transverse acceleration
\begin{equation}
\dot{v}_T=-\frac{\alpha Z}{E}\frac{b_0}{\sqrt{t^2+b_0^2+R^2}^3},
\end{equation}
and
\begin{displaymath}
v_T(t)=-\frac{\alpha Z}{E}
\frac{b_0}{b_0^2+R^2} \frac{t+\sqrt{t^2+b_0^2+R^2}}
{\sqrt{t^2+b_0^2+R^2}}=
\end{displaymath}
\begin{equation}
-\frac{\alpha Z}{E} \frac{b_0}{b_0^2+R^2} \frac{t+\sqrt{r^2+R^2}}
{r^2+R^2}. \label{velocity}
\end{equation}
From $v_T(t \rightarrow \infty)=-\frac{2\alpha Z}{E}\frac{b_0}
{b_0^2+R^2}$ we obtain for a pure Coulomb field $(R \ll b_0)$
the well-known transverse momentum transfer
\begin{equation}
\Delta k_T = \frac{2 \alpha Z}{b_0}.
\end{equation}
Furthermore, we obtain from (\ref{velocity})
\begin{equation}
b(z)=b_0-\frac{\alpha Z}{E}
\frac{b_0 (z+\sqrt{r^2+R^2})}{b_0^2+R^2},
\end{equation}
The focusing factor is then given by $(\beta=\alpha Z/E)$
\begin{displaymath}
f^{-1}=\Bigl( 1-\frac{z+\sqrt{r^2+R^2}}{b_0^2+R^2} \beta \Bigr) \times
\end{displaymath}
\begin{equation}
\Bigl(1- \Bigl( \frac{z+\sqrt{r^2+R^2}}{b_0^2+R^2}-
\frac{2 b_0^2(z+\sqrt{r^2+R^2})}{(b_0^2+R^2)^2}+
\frac{b_0^2}{(b_0^2+R^2)\sqrt{r^2+R^2}}
\Bigr) \beta \Bigr). \label{focus2}
\end{equation}
Since our calculation is not exact, we keep
only the relevant first order in $\beta$ in (\ref{focus2}).
For the focusing factor in the center of the nucleus follows
\begin{equation}
f(0)=\Bigl(1-\frac{\beta}{R} \Bigr)^{-2} \sim \Bigl(1-\frac{V(0)}{E}
\Bigr)^2 \sim (k'_i/k_i)^2, \label{classicalfocus}
\end{equation}
i.e. one obtains the focusing factor used in the effective
momentum approximation, where the increased density near the
nucleus is taken into account by multiplying the wave
function by a suitable factor $\sim f^{1/2}$.
In (\ref{classicalfocus}), $k_i=E/c$ denotes the asymptotic momentum
of the highly relativistic incident particles,
whereas $k'_i$ is the momentum of
a particle with impact parameter $b_0=0$ in the center
of the nucleus, which is given
by $k'_i=k_i+\alpha Z/(Rc)$. c is the speed of light.
The amplitude of the
wave function is then correctly normalized in the region of the
nucleus, where the nucleon knockout process is taking place.
The discussion given above
is only classical, but it is sufficient to show the
most important features of the focusing effect.
E.g., from (\ref{focus2}) is it obvious that the amplitude of the
wave function continues to increase on the rear side of
the nucleus due to the deflection of the incoming particle wave.

Knoll \cite{Knoll} derived the focusing effect from a high energy
partial wave expansion, following previous results given
by Lenz and Rosenfelder \cite{Lenz,Rosenfelder}.
For the incoming particle wave expanded around the center of
the nucleus he obtained:
\begin{displaymath}
\chi_k^+=e^{i \delta_+} (k'/k) e^{i \vec{k'}\vec{r}} \times
\end{displaymath}
\begin{equation}
\{1+a_1r^2-2a_2\vec{k'}\vec{r}+ia_1r^2 \vec{k'}\vec{r}+ia_2[
(\vec{k'} \times \vec{r})^2+
\vec{\sigma} (\vec{k'} \times \vec{r})] \} u_k, \label{expknoll}
\end{equation}
where $\vec{k'}$ is parallel to $\vec{k}$, and where $\sigma$
acts on the spinor $u_k$ to describe spin dependent
effects, which are very small in our cases of interest.
An analogous equation holds for the distortion of the outgoing wave.
The parameters $a_{1,2}$ depend on the shape of the
potential. For a homogeneously charged sphere with radius $R_s$
they are given by
\begin{equation}
a_1=-\frac{\alpha Z}{6 k' R_s^3} , \quad 
a_2=-\frac{3 \alpha Z}{4 k'^2 R_s^2} \label{para} .
\end{equation}
The increase of the amplitude of the wave while passing
through the nucleus is given by the $-2a_2\vec{k'}\vec{r}$-term.
Taking our classical result for a screened potential leads to
$a_2 \sim \alpha Z/ k'^2 R^2$, a result which
cannot be compared directly to (\ref{para}) due to the different
shape of the potentials.
Due to the fact that the amplitude of the wave function is
smaller in the upstream side of the nucleus but larger
by a similar amount on
the downstream side, the influence of the $-2a_2\vec{k'}\vec{r}$-term
on the cross section in general is not very large.
Minor effects can be observed for large scattering angles.
Also the spin related term is of minor importance for
highly relativistic electron energies.

The $a_1 r^2$-term accounts for the decrease of the focusing
also in transverse direction. For the cases of interest in this
paper, it leads to 1-2 percent effects in the cross sections.
Imaginary terms like $ia_2
(\vec{k'} \times \vec{r})^2$ describe the deformation of the
wave front near the center of the nucleus. They could be obtained
correspondingly by an expansion of the eikonal phase in that region,
and $\delta_+$ is the eikonal phase in the center of the nucleus.

In the present work, the eikonal phase is obtained directly
from an analytic expression as described in Sect. 2.2,
whereas the focusing is calculated using Knoll's results given
above for a homogeneously charged sphere.

\subsection{Electrostatic potential of the nucleus}
For our present eikonal calculations, we use a potential energy of the
electron of the form
\begin{equation}
(\alpha Z)^{-1}V(r)=-\frac{r^2+\frac{3}{2}R^2}{(r^2+R^2)^{3/2}}
-\frac{24}{25 \pi} \frac{R^2 R' r^4}{(r^2+R'^2)^4} \, ,
\label{potential}
\end{equation}
which goes over into a Coulomb potential for $r \rightarrow 
\infty$, and which is a good approximation for the potential generated
by the Woods-Saxon-like charge distribution of a nucleus (see Fig.
{\ref{figpot}}). $R'$ can be used as an additional fit parameter. A good choice is $R'=0.5174 R_e$.
Furthermore, expression (\ref{potential}) has the
advantage that it is possible to derive an analytic
expression for the eikonal phase, which is convenient
when the eikonal phase has to be calculated numerically in
a computer program.

The eikonal phase turns out
to be divergent for a Coulomb-like potential,
but it is possible to
regularize the eikonal phase by subtracting a screening potential
$\sim (r^2+a^2)^{-1/2}$ with $a \gg R$ from (\ref{potential}),
such that the potential
falls off like $r^{-2}$ for large $r$.
The divergence can then be absorbed
in a constant divergent phase $\sim \log(a)$ without
physical significance, when the limit $a \rightarrow
\infty$ is taken. It is quite instructive to calculate the
eikonal phase for the simple screened potential \cite{Glauber}
\begin{equation}
V_1(r)=-\frac{\alpha Z }{\sqrt{r^2+R^2}} \, , \quad
V_1^a(r)=-\frac{\alpha Z }{\sqrt{r^2+R^2}}+
\frac{\alpha Z }{\sqrt{r^2+a^2}} \, .
\end{equation}
One obtains for a particle incident parallel to the z-axis
for impact parameter $b$ ($r'^2=b^2+z'^2$)
\begin{displaymath}
\chi_1^a=\alpha Z \int_{-\infty}^{z} dz' \,
\Bigl( \frac{1}{\sqrt{r'^2+R^2}}-
\frac{1}{\sqrt{r'^2+a^2}} \Bigr)=
\end{displaymath}
\begin{equation}
\alpha Z \log \frac{(z+\sqrt{r^2+R^2})(b^2+a^2)}
{(z+\sqrt{r^2+a^2})(b^2+R^2)},
\end{equation}
where $r^2=b^2+z^2$, and therefore for the regularized
eikonal phase
\begin{equation}
\chi_1=\lim_{a \to \infty} (\chi_1^a - \alpha Z \log(a))=
\alpha Z \log \Bigl( \frac{z+\sqrt{r^2+R^2}}{b^2+R^2} \Bigr) .
\label{regularizedeikonal}
\end{equation}
which is defined up to a constant phase.
Taking the gradient of $\chi_1$ in transverse direction
\begin{equation}
\frac{\partial \chi_1}{\partial b}=-\frac{\alpha Z b}{b^2+R^2} \frac{z+\sqrt{r^2+R^2}}
{r^2+R^2},
\end{equation}
gives for the transverse momentum transfer
the same result as the classical expression (\ref{velocity}).
This illustrates the fact that the eikonal approximation also
accounts for the transverse modification of the particle momentum.
The charge density
\begin{equation}
\rho(r)=-\frac{1}{e r} \partial^2_r (r V(r))
\end{equation}
corresponding to the potential given by Eq. (\ref{potential})
satisfies
\begin{equation}
\langle \rho \rangle = eZ \, ,
\end{equation}
\begin{equation}
\langle r^2 \rho \rangle = \frac{3}{5}R^2 eZ \, ,
\end{equation}
i.e. we can indeed identify $R^2$ with the equivalent radius of a
homogeneously charged sphere $R_e^2$
which is given approximately by
\begin{equation}
R_e=1.128 A^{1/3}\, \mbox{fm}+
2.24 A^{-1/3} \, \mbox{fm}
\end{equation}
for nuclei with $A>20$. $R_e$ can be related to the rms radius
$R_m$ by
\begin{equation}
R_m^2=\frac{3}{5} R_e^2 \, .
\end{equation}

For a simple potential $\sim (r^2+R^2)^{-1/2}$, the
rms radius does not exist, since the corresponding
charge distribution does not fall off fast enough.

The expressions necessary for the calculation of the
eikonal phase for the potential (\ref{potential})
are given in the appendix.

\begin{figure}
        \centering
        \includegraphics[width=9cm]{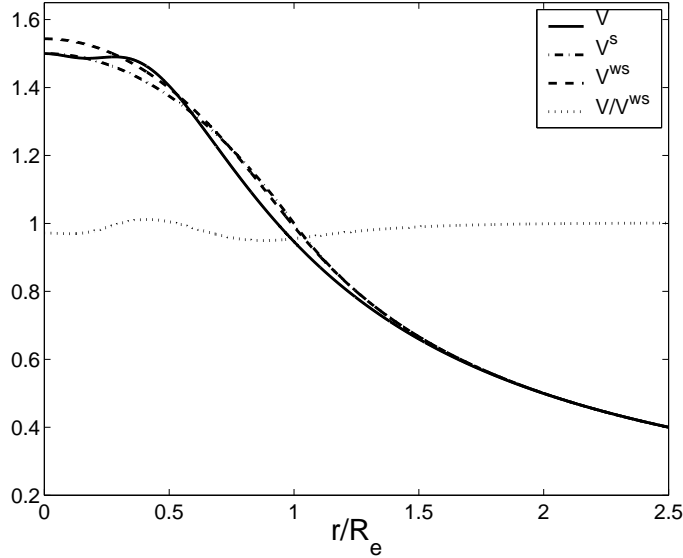}
        \caption{Comparison of different model potentials: Potential
        $V$ used in our calculations, potential $V^s$ generated by
        a homogeneously charged sphere with radius $R_e$ and potential
        generated by a corresponding Woods-Saxon charge distribution.}
        \label{figpot}
\end{figure}

\subsection{Scattering cross section}
For the sake of notational simplicity and in order to give
a transparent introduction to the method used for the calculation
of the CC,
we will neglect the dependence of the interactions on the spin and
internal structure of the particles.
Therefore, the particle wave functions are replaced
by scalar fields in the following discussion.
We call the positively charged particle the 'proton',
whereas the negatively charged particle is the 'electron'.
The simplified scalar expression for the electron current
can then be replaced by the correct Dirac current (\ref{current1}) in
a straightforward way. The nucleon current which we used
for our calculations is discussed below.

The lowest order transfer current of an electron with initial/final
state wave function $\phi_{e}^{i,f}$ and a scalar (point-like) proton
with initial/final state wave function $\phi_{p}^{i,f}$ is given by
\begin{equation}
j_{e}^\mu=+ie(\phi_{e}^{f \dagger} \partial^\mu \phi_{e}^i-
\phi_{e}^i \partial^\mu \phi_{e}^{f \dagger}) \, ,
\end{equation}
\begin{equation}
j_{p}^\mu=-ie(\phi_{p}^{f \dagger} \partial^\mu \phi_{p}^i-
\phi_{p}^i \partial^\mu \phi_{p}^{f \dagger}).
\end{equation}
For the scattering cross section
of an electron with initial and final
momentum $k_i^\mu=(\epsilon_i,\vec{k}_i)$, $k_f^\mu=
(\epsilon_f,\vec{k}_f)$ off a proton with momenta
$p_i^\mu=(E_i,\vec{p}_i)$, $p_f^\mu=(E_f,\vec{p}_f)$
we obtain in Born approximation
the second order contribution in the coupling constant $e$
to the differential cross section
($e^2=4 \pi \alpha$):
\begin{displaymath}
d \sigma = \frac{1}{4 \epsilon_i E_i |v^e_i-v^p_i|}
\Bigl( \frac{d^3k_f}{2 \epsilon_f (2 \pi)^3}
\frac{d^3 p_f}{2E_f(2 \pi)^3} \Bigr)
\end{displaymath}
\begin{equation}
\times (2 \pi)^4 \delta^{(4)}(k_i+p_i-k_f-p_f) | M |^2 ,
\end{equation}
where the matrix element $M$ is given by the product of the
currents and the photon propagator
\begin{equation}
M = e^2 \frac{(k_i+k_f)_\mu (p_i+p_f)^\mu}{(k_i-k_f)^2} .
\end{equation}
The corresponding expression for the physical situation where electrons
and nucleons are treated as particles with spin can be found,
e.g., in \cite{Udias2}.

Since we consider proton knockout by highly relativistic
electrons from a nucleus at rest,
the velocity in the flux term $|v^e_i-v^p_i|$
will be set to the speed of light.

If the initial proton is in a bound state with wave function
\begin{equation}
\phi_{p}^i(\vec{r},t)=(2 \pi)^{-3/2} \int d^3 q \, 
\hat{\psi}(\vec{q} \,)e^{i (\vec{q} \vec{r}-E_i t)},
\end{equation}
the transition probability is replaced by $(p_i^\mu=(E_i,\vec{q}\,))$
\begin{displaymath}
(2 \pi)^4 \delta^{(4)}(k_i+p_i-k_f-p_f) | M |^2 \rightarrow
\end{displaymath}
\begin{displaymath}
2 \pi \delta(\epsilon_i+E_i-\epsilon_f-E_f)
\end{displaymath}
\begin{displaymath}
\times \Bigl| \, e^2
\int d^3 q \int d^3 r \, \frac{(k_i+k_f)_\mu (p_i+p_f)^\mu}
{(E_i-E_f)^2-(\vec{q}-\vec{p}_f)^2}
\, \hat{\psi}(\vec{q}) \,
e^{i(\vec{k}_i+\vec{q}-\vec{k}_f-\vec{p}_f) \vec{r}} \, \Bigl|^2
\rightarrow
\end{displaymath}
\begin{displaymath}
(2 \pi)^4 \delta(\epsilon_i+E_i-\epsilon_f-E_f)
\end{displaymath}
\begin{equation}
\times \Bigl| \, e^2
\frac{(k_i+k_f)_\mu (2p_f+k_f-k_i)^\mu}{(k_i-k_f)^2} 
\hat{\psi}(\vec{k_f}+\vec{p_f}-\vec{k_i}) \, \Bigr|^2 . \label{born}
\end{equation}

In the distorted wave Born approximation, the presence of
the Coulomb field is taken into account by using exact
wave functions of the particles in the Coulomb
field. In the eDWBA, these wave functions are replaced by their
eikonal approximation. Therefore, the electron current is
modified by the eikonal phase and due to the focusing of the
electron wave.
For electrons with potential energy $V(r)$ in the central
Coulomb field of the nucleus we must replace
\begin{equation}
\int d^3 q \int d^3 r \, \frac{(k_i+k_f)_\mu (p_i+p_f)^\mu}
{(E_i-E_f)^2-(\vec{q}-\vec{p}_f)^2)}
\, \hat{\psi}(\vec{q}) \,
e^{i(\vec{k}_i+\vec{q}-\vec{k}_f-\vec{p}_f) \vec{r}}
\end{equation}
by
\begin{equation}
\int d^3 q \int d^3 r \, \frac{(\tilde{k}_i+\tilde{k}_f)_\mu (p_i+p_f)^\mu}
{(E_i-E_f)^2-(\vec{q}-\vec{p}_f)^2)}
\, \hat{\psi}(\vec{q}) \, f(\vec{r})
e^{i(\vec{k}_i+\vec{q}-\vec{k}_f-\vec{p}_f) \vec{r}+i \chi(\vec{r})},
\label{inteiko}
\end{equation}
where
\begin{equation}
\tilde{k}_i^\mu=(\epsilon_i-V(r),\vec{k}_i+\vec{\nabla} \chi_1) \, ,
\end{equation}
\begin{equation}
\tilde{k}_f^\mu=(\epsilon_f-V(r),\vec{k}_f-\vec{\nabla} \chi_2)
\end{equation}
correspond to the modified electron current in the eikonal
approximation, and $f(\vec{r})$ denotes the focusing factor
resulting from the incoming and outgoing particle wave function.
The gradient and potential terms in the electron current are an artefact
of our treatment of electrons as scalar particles. They appear
in a similar fashion in the Gordon form of the electron current.
The standard Dirac form of the current
\begin{equation}
j^\mu=e \bar{\Psi} \gamma^\mu \Psi, \label{current1}
\end{equation}
can be split into a convective current and a spin current
after some algebra by making use of the Dirac equation. One obtains
\begin{equation}
j^\mu=\frac{ie}{2m} \bigl[ \bar{\Psi} \partial^\mu \Psi -(\partial^\mu
\bar{\Psi}) \Psi +\frac{2ie}{m} \bar{\Psi}\Psi A^\mu
\bigl]
+\frac{e}{2m} \partial_\mu [\bar{\Psi} \sigma^{\mu \nu} \Psi ] \, .
\label{current2}
\end{equation}
The convective and the spin current are separately conserved
and gauge invariant.
But it is not advisable to use the Gordon form (\ref{current2})
in conjunction with the inexact eikonal approximation of the electron
wave function, because the electron current is then described less
accurately than by the standard expression (\ref{current1}).

It is now no longer possible 
to evaluate the coordinate space integral in a trivial
way. As a first approximation we assume that the initial momentum
("missing momentum") of the bound proton is given approximately
by the external momenta, i.e.
\begin{equation}
\vec{q}=\vec{k}_f+\vec{p}_f-\vec{k}_i-\Delta \vec{q}, 
\quad \Delta \vec{q} \sim 0.
\end{equation}
Then we can expand
\begin{displaymath}
\Bigl((\epsilon_i-\epsilon_f)^2-(\vec{q}-\vec{p}_f)^2\Bigr)^{-1}=
\end{displaymath}
\begin{displaymath}
\Bigl((\epsilon_i-\epsilon_f)^2-(\vec{k}_i-\vec{k}_f)^2
-2(\vec{k}_i-\vec{k}_f) \Delta \vec{q} -\Delta \vec{q}^{\, 2}
\, \Bigr)^{-1}=
\end{displaymath}
\begin{equation}
-\frac{1}{Q^2} \Biggl(1-\frac{2(\vec{k}_i-\vec{k}_f)\Delta \vec{q}
+\Delta \vec{q}^{\, 2}}
{Q^2}+\frac{4[(\vec{k}_i-\vec{k}_f)\Delta \vec{q} \, ]^2}
{Q^4}+ \ldots \Biggr) \, , \label{expansion}
\end{equation}
where $Q^2=(\vec{k}_i-\vec{k}_f)^2-(\epsilon_i-\epsilon_f)^2$.
The zeroth order contribution to (\ref{inteiko}) is then given by the term
\begin{equation}
-\frac{(2 \pi)^{3/2}}{Q^2}
\int d^3 r \, (\tilde{k}_i+\tilde{k}_f)_\mu
(p_i+p_f)^\mu \, \psi(\vec{r}) \, f(\vec{r})
e^{i(\vec{k}_i-\vec{k}_f-\vec{p}_f) \vec{r}+i \chi(\vec{r})}.
\end{equation}
The $\Delta \, \vec{q}$-terms stemming from the
expansion above and from the expression for the proton
current act then as a gradients
on the eikonal phase in real space according to
\begin{displaymath}
\int d^3 q \int d^3 r \, \Delta \vec{q} \, \hat{\psi}(\vec{q}) 
e^{i(\vec{k}_i+\vec{q}-\vec{k}_f-\vec{p}_f) \vec{r}+i \chi(\vec{r})}=
\end{displaymath}
\begin{equation}
-(2 \pi)^{3/2} i \int d^3 r \, \psi(\vec{r}) 
e^{i(\vec{k}_i-\vec{k}_f-\vec{p}_f) \vec{r}}
\vec{\nabla} e^{i \chi(\vec{r})},
\end{equation}
such that we obtain from (\ref{expansion}) an expansion of the
integral (\ref{inteiko}) which contains higher derivatives of the
eikonal phase. For the actual calculations, it proved sufficient
to include all terms containing first and second order
derivatives. This way, the matrix element (\ref{inteiko})
can be reduced to three-dimensional integrals which are
numerically tractable. This is due to the fact that the integrals extend
only over a finite volume around the nucleus. 
For the case presented in Fig. 4,
second order derivative terms contribute about one percent
to the total cross section.

\subsection{Nucleon-nucleus interaction}
As a first step, we adopted also an eikonal approximation for
the proton wave function, in order to improve the poor approximation
of describing the outgoing proton by a plane wave.
The corresponding treatment can be carried out
along the same lines as discussed above.
In order to be as realistic as possible,
we use an energy-dependent volume-central part of an
optical model potential as given in a recent work \cite{Koning}.
The potential is given as the sum of a Woods-Saxon potential and
the Coulomb potential of the nucleus for protons.
The depth of the Woods-Saxon part of the potential
depends on the energy $E$ of the proton (in MeV) by
\begin{equation}
V_{WS}(E)=v_1 [ 1- v_2 (E-E_f)+v_3 (E-E_f)^2-v_4(E-E_f)^3 ] \, ,
\end{equation}
where
\begin{equation}
v_{1}=67.2 , \, v_2=7.9 \cdot 10^{-3}, \,
v_3=2.0 \cdot 10^{-5}, \, v_4=7 \cdot 10^{-9}, \, E_f=-5.9.
\end{equation}
For neutrons we have
\begin{equation}
v_{1}=50.6, \, v_2=6.9 \cdot 10^{-3}, \,
v_3=1.5 \cdot 10^{-5}, \, v_4=7 \cdot 10^{-9}, \, E_f=-5.65.
\end{equation}
It has been noted in \cite{Jin3} that when comparing the equivalent
central potentials generated by the sum of the scalar and vector
part of relativistic potentials, the real part of the optical
potential is $10$ MeV deeper for $^{40}$Ca. A comparison of a
standard nonrelativistic central potential and a corresponding potential
obtained from a Dirac equation based procedure given in
\cite{Udias} shows good agreement for $^{208}$Pb (the difference is
of the order of only 2 MeV), as well as a good agreement with
the optical model potential given above.

The imaginary part of the optical potential was not taken
into account in our calculations.
The imaginary part is intended to describe the loss of flux
in proton (neutron) elastic scattering. In inclusive processes,
only the electrons are observed, and one does not need to take into account
whether or not the ejected nucleons got "lost", i.e. initiated
some subsequent nuclear reaction.

The same wave functions for the outgoing nucleon are used
for the PWBA and eDWBA calculation,
since we are only interested in the electronic part of CC.
The eikonal approximation, when also used
for the calculation of the wave function of the outgoing nucleons,
leads to a considerable additional calculational effort.
We found that an effective momentum
approximation for the proton calculated from the energy-dependent
optical model potential leads to equally good results,
and reduces the calculational effort considerably.
It is obvious that the plane waves used as final states
are not orthogonal to the initial bound states,
but the results obtained for the cross sections are still
quite satisfactory for our purposes despite of the fact that
we use a single particle and a plane wave approximation.
We studied also the inclusion of a focusing factor for the proton
wave function in order to improve upon the plane wave approximation.
Although such a procedure might be used to improve the theoretical
cross sections artificially, it is only an ad hoc procedure
and was therefore not used for the calculations presented in this paper.

In a first attempt to calculate CC for a heavy nucleus like
$^{208}$Pb, we used the harmonic oscillator shell model wave functions
for the bound protons. These wave functions
are easily accessible for numerical
calculations due to their simple analytic form. Harmonic
oscillator wave functions are satisfactory approximations
for nucleon wave functions especially when
they are scaled according to the root mean square (rms) radius of the
individual shells;
due to the requirement of orthogonality of initial states,
the same scaling must be applied to all wave functions.
The contribution of partially filled upper shells to the inclusive
proton knockout cross section was taken into account by a weighting
factor for the total shell contribution according to the number of
occupied states.

As an alternative we also used wave functions
generated from a Woods-Saxon potential
for the protons.
The Woods-Saxon potential, which
included an LS coupling term, was optimized in
such a way that the experimental binding energies
of the upper proton shells and the rms of the
nuclear charge distribution were reproduced correctly.
A comparison of calculated results
showed that the ratio $\sigma_{PWBA}/\sigma_{CC}$
for a kinematic situation as presented in Fig. 4 differs
only by about one percent for the two models.
A comparison of the charge distribution in a Pb nucleus
resulting from the two models is shown in Fig. {\ref{figcharge}}.
The harmonic oscillator model overestimates the charge
density in the center of the nucleus, but the impact of this
mismatch is reduced due to the small volume of the
central region.
The rms radius of the neutron density distribution was taken from a
recent study presented in \cite{Clark}.
\begin{figure}
        \centering
        \includegraphics[width=9cm]{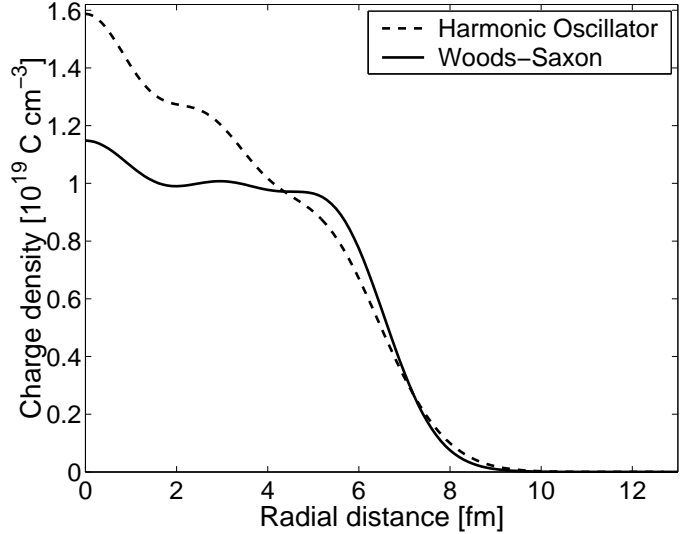}
        \caption{Charge density of the Pb nucleus derived from 
        the  harmonic oscillator model and a Wood-Saxon potential fit.}
        \label{figcharge}
\end{figure}
The binding energies, which also enter in the calculation of the
wave functions of the outgoing protons and in the phase space
factors, were taken from \cite{Batenburg}.

The current of a spinless proton could be modeled by the widely used 
electric Sachs form factor $G_E$ \cite{Hoehler}:
\begin{equation}
G_E(Q^2)=(1+Q^2/(843 \, \mbox{MeV})^2)^{-2} \, , \label{form}
\end{equation}
which can be expanded by the same method as the photon propagator
(\ref{expansion}). It is expected that
(\ref{form}) also provides a good description of the
form factor of a proton inside a nucleus \cite{Jourdan96a}.
For the energy range considered in this paper, the
contribution from magnetic scattering must not be neglected.
For the calculations, we used the free nucleon current
given in \cite{Jin1} and introduced by de Forest \cite{deForest}
\begin{equation}
j^\mu_n=e\bar{\Psi}_{n,f} {\hat{J}}_n^\mu \Psi_{n,i} \, ,
\end{equation}
where
\begin{equation}
{\hat{J}}_n^\mu(cc2)
=F_1 \gamma^\mu + \frac{i \kappa}{2 m_n} \sigma^{\mu \nu} q_\nu F_2 ,
\end{equation}
and $q_\nu=p'_\nu-p_\nu$ denotes here the difference between the
proton four-momenta of the initial and final state,
$\kappa$ the anomalous magnetic moments of the corresponding nucleon.
In configuration space, the three-momenta have to be considered as
gradient operators.
$F_{1,2}$ are the nucleon form factors related in the usual way
to the electric and magnetic Sachs form factors.

The relativistic form of the
wave function for the bound protons and neutrons
was constructed from the nonrelativistic wave function. Whereas
the large component of the corresponding correctly normalized
Dirac spinor was obtained directly from the nonrelativistic
wave function, the small component was derived from the large component
in a straightforward way by treating the nucleon wave
function as a free Dirac plane wave.
The momenta of the nucleons which were used in the current
can be calculated from energy and momentum conservation.
But in order to improve our model for the current, we adopted additionally
an effective momentum approximation for the nucleons, where the influence
of the energy-dependent optical potential on the ejected nucleons
was taken into account.
We found that different choices for the nucleon current
do not affect significantly the ratio of the PWBA and eDWBA calculations.
This is not the case for the cross section
themselves. An alternative choice to the cc2 current
is the cc1 current given by the operator \cite{deForest}
\begin{equation}
{\hat{J}}_n^\mu(cc1)=(F_1+\kappa F_2) \gamma^\mu
-\frac{({p'}^\mu+p^\mu)}{2 m_N} \kappa F_2,
\end{equation}
which was also used in order to check how the ratio of the cross
sections behaves for different models of the nuclear current.
None of the expressions for the current (cc1,cc2)
is fully satisfactory and both expressions fail to fulfill current
conservation, but given the fact that we focus mainly on the electronic
part of the problem, the simple choices given above provide a satisfactory
description of the proton current. For the cc2 choice, integration
by parts allows one in a simple way to get rid of the momentum
operators acting on the nucleon wave functions.

Some critical remarks which highlight the approximations made
within our single particle model are in order.
Due to the fact that we are working in real space,
we scaled the nucleon wave functions according to the experimental
rms radius of the nucleus. This does not automatically imply
that the momentum distribution of the nucleons is described
very accurately. The difference between the theoretical Fermi
momentum obtained in our model and the experimental value is of
the order of 10\%.
Additionally, the choice to use an effective momentum
approximation for the nucleons is an {\em{ad hoc}} prescription, which
differs from the naive plane wave approximation for
nucleons where the influence of an optical
potential is neglected. In both cases, one has to accept that
focusing effects of the nucleon wave functions
in the nuclear region are missing and furthermore that
unitarity is violated to a certain degree.

Kim {\em{et al.}} \cite{Jin1} presented calculations where
plane waves and Dirac wave functions were used for the
electrons, but the nucleon wave functions were calculated
in both cases from a relativistic $\sigma-\omega$ model.
For the kinematical situation presented in Fig. 6 in the
next section, their theoretical cross sections
are slightly below the experimental values, whereas our model
leads to cross sections slightly above the experimental values.
The advantage of our model is that we obtain a curve which
shows a very similar behavior as the experimental curve, although this
should not be considered as a virtue of our method.
We expect that our model provides reliable results only for
the {\em{ratio}} of cross section (were one uses plane waves or
'exact' wave functions for the electrons), which can be used
for the analysis of experimental data. We checked this assumption
by varying the optical potential, the binding energies and
by rescaling the wave functions within reasonable limits.
The relatively large impact of such variations on the cross section
is divided out for the most part in the ratio of the cross
sections.

Kim {\em{et al.}} \cite{Resler} presented also a simplified
model where free plane wave functions for the ejected nucleons
and harmonic wave functions for the bound nucleons were used.
Adopting the same strategy in our case, we obtain nearly
identical results as those presented in \cite{Resler} when
we scale our wave functions according to the experimental
value of the nuclear Fermi momentum.

The numerical evaluation of (\ref{inteiko}) was performed
by putting the nucleus on
a three dimensional grid with a side length of 36 fm
and using the Simpson method as a very simple
but efficient integration tool.
The number of necessary grid points is mainly dictated by
the wave length of the oscillatory term
$e^{i(\vec{k}_i-\vec{k}_f-\vec{p}_f)}$
and was in the range of $60^3$ to $120^3$ in order
to ensure an accuracy of $10^{-2}$ percent for the values of
the integrals.

It is instructive to have a (classical) look a the approximate
size of the different effects
which lead altogether to the CC. As a specific case we choose
the initial energy of the electron $\epsilon_i=485$ MeV,
scattering angle $\theta_e=60^o$ and and energy transfer
$\omega=100$ MeV. Additionaly, we take the Coulomb
potential energy $V(0)$ of an electron
in the center of a $^{208}$Pb nucleus as
$-25$ MeV. The focusing then enters in the cross section as a factor
$(510/485)^2(410/385)^2 \sim 1.25$, i.e. it leads
to an increase of the cross section of about $25$\%.
From the photon propagator term $\sim q^{-4}$ on the contrary
we obtain a reduction
by a factor of $0.80$. If we use the potential
at the surface of the nucleus $-17$ MeV, the reduction factor is
$0.85$. This discrepancy
illustrates the importance of having an accurate
description of the electron wave function.
Furthermore, the form factor of the proton enters the
cross section as a factor of the order of $\sim (1+Q^2/(843 \, \mbox{MeV})^2)^{-4}$,
reducing the cross section by $6$-$9$\% (for V(0)=
$-17$ Mev to $-25$ MeV).
Finally, the nontrivial change of the interaction which involves
the detailed structure of the current via the nucleon wave functions
gives an effect of several percent. All these effects described above
combine to a CC which is almost zero for the present example.

\section{Results}
\subsection{Comparison of the results to other approaches}
Fig. 4 shows the ratio of the inclusive cross section for
nucleon knockout with and without CC
for electrons with initial energy of $\epsilon_i=485$ MeV
and scattering angle $\Theta_e=60^o$ on $^{208}$Pb.
Neutron knockout contributes by about
20\% to the total nucleon knockout cross section in the considered kinematical region.
The results obtained by Kim {\em{et al.}} \cite{Jin1}
and the eikonal approximation
are in good agreement, also for different kinematics as
e.g. shown in Fig. 5.
The result for the local effective momentum approximation
(LEMA) of Kim {\em{et al.}}, which is based on parameters
that were fitted to the exact calculation,
is also fairly close.

The conditions presented in Fig. 5 correspond to a similar momentum
transfer, but large scattering angle $\Theta_e=143^o$
and incident energy $310$ MeV.
The initial electron energy is smaller,
and the effect of CC is more pronounced.
CC lead to a sizeable modification
in the longitudinal and transverse
responses that can be extracted from the data via Rosenbluth plots.
The linearity of the Rosenbluth plots served also as an independent
check for the validity of our plane wave calculations.
\begin{figure}
        \centering
        \includegraphics[width=9cm]{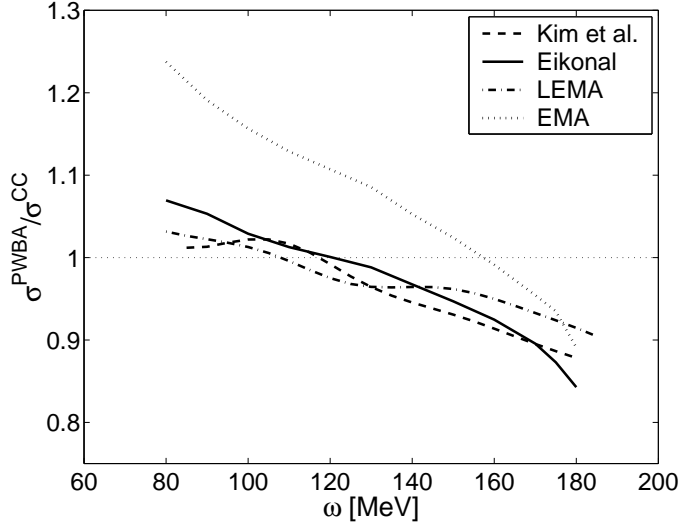}
        \caption{Comparison of the Coulomb corrections
        for different approaches. The dotted curve shows the
        ratio $\sigma_{PWBA}/\sigma_{EMA}$ according
        to the model used in this paper.}
        \label{fig4}
\end{figure}
\begin{figure}
        \centering
        \includegraphics[width=9cm]{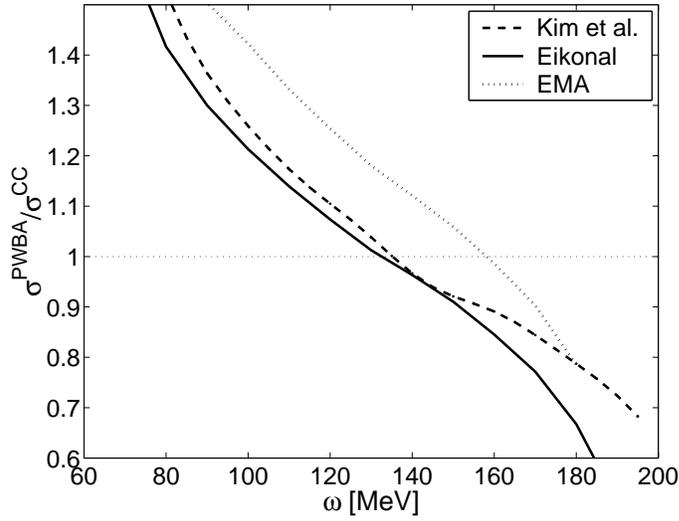}
        \caption{$\sigma_{_{PWBA}}/\sigma_{_{CC}}$ for $\epsilon_i
        =310$ MeV and $\Theta_e=143^o$. Dotted curve:
        $\sigma_{PWBA}/\sigma_{EMA}$ as in Fig. 4.}
        \label{fig5}
\end{figure}

The calculations clearly indicate that the effective
momentum approximation, which leads to admissible results for
light nuclei where the CC are relatively small,
is useless for highly charged nuclei.
The curve in Fig. 5 was calculated by
choosing a $\Delta k=25$ MeV/c according to Eq. (\ref{deltaeff}).
Taking smaller values for $\Delta k$ corresponding to the electrostatic
potential at the surface of the nucleus does not change the situation
significantly.

It is important to take into account the final state interaction
of the proton by an energy-dependent optical potential.
For protons with energies above $100$ MeV,
the potential is becoming increasingly repulsive.
Calculations for $\omega > 180$ MeV in our single particle shell
model are not applicable,
since pion production is not included in our calculations, and
correlation effects are becoming increasingly important
for large $\omega$.
Results for $\omega < 80$ MeV may also become dubious due to
the large distortion of the outgoing proton wave function
by the final state interaction,
but the simple approximation given in this work allows
for a good estimation of CC in a relatively wide range
of energies.

\begin{figure}
\begin{center}
        \includegraphics[width=9cm]{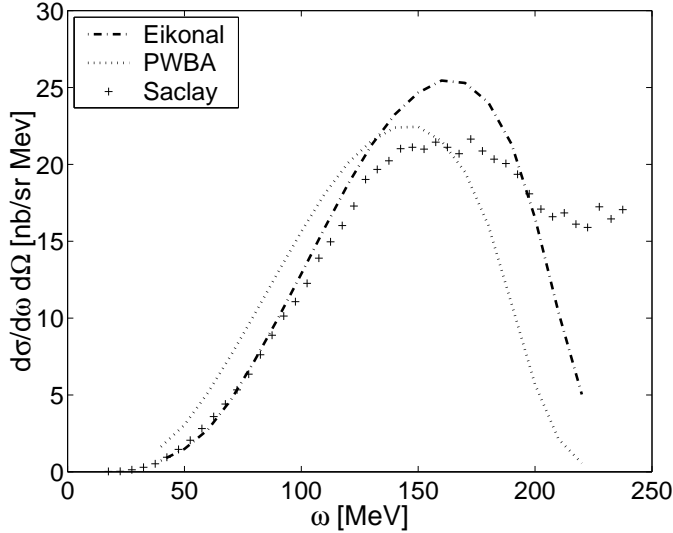}
        \caption{Comparison of theoretical cross sections obtained
                 in PWBA and eikonal approximation to experimental
                 Saclay data ($\epsilon_i=310$ MeV and $\Theta_e=143^o)$.}
        \label{fig6}
\end{center}
\end{figure}

It is a well-known phenomenon that plane wave calculations
have a tendency to overestimate cross sections in the region of
low electron energy loss,
whereas cross sections are underestimated in the
high energy loss region. The reason for this is the fact that
the use of plane waves for the ejected nucleons is clearly a rough
approximation, and the generally used single particle
model does not contain
contributions arising from correlations and meson-exchange currents.
Inelastic scattering from the nucleons, i.e. excitation of
the delta resonance, is also absent.

Nevertheless, the agreement between
experimental data taken at Saclay
\cite{Zghiche} and our eikonal corrected calculations
is quite satisfactory, as displayed in {Fig. 6}
in order to give a typical example.
Like in the case $\Theta_e=60^o$,
one observes that the CC become very large for high energy
transfer $\omega \simeq 180$ MeV. This result should be handled
with care, since the calculated cross sections become small in this
energy region.
The electroproduction of the delta resonance will have to be
included to more reliably estimate the CC at very large energy loss.

Overall, we find that the eikonal approach gives results
that are very close to the results from exact calculations.
This approach can thus be used in practical applications where
one greatly benefits from the much lower calculational effort involved in
the eikonal description. At the same time, our calculation shows again
that the often-used effective momentum approximation is inadequate for
a quantitative treatment of Coulomb corrections.

\subsection{Study of the Coulomb distortion by comparing
quasielastic electron and positron scattering}
The comparison of experimentally determined quasielastic
cross sections induced with electrons and with positrons represent
an ideal test for any theoretical approach to quantitatively
describe Coulomb distortions.
Here we consider a recent experiment by Gu\`eye {\em{et al.}}
\cite{Gueye} which was used to give an experimental proof of
the validity of the EMA for inclusive quasielastic electron
scattering. The experiment by Gu\`eye {\em{et al.}}
compared quasielastic scattering of electrons
on $^{208}$Pb with quasielastic scattering of positrons
at the same kinematics. 

\begin{figure}[htb]
\begin{center}
        \includegraphics[width=9cm]{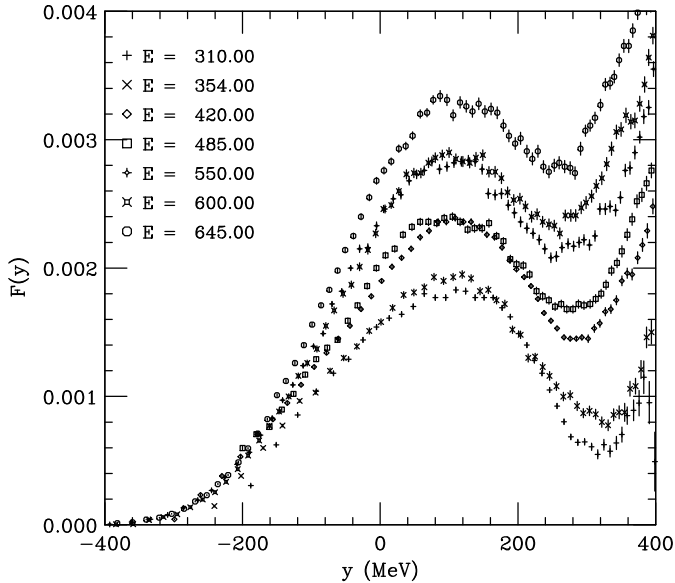}
        \caption{Scaling function $F(y)$ obtained
         from cross sections measured by Zghiche
         {\em{et al.}} on $^{208}$Pb at an electron scattering angle of
         60 degrees and initial electron energy $E$.
         $F(y)$ is obtained from the cross sections by dividing
         out the Mott cross section and the $q$-dependence
         of the nucleon form factors (both smooth functions
         of incident energy).}
        \label{fig7}
\end{center}
\end{figure}
With the assumption that the EMA is a valid approach the authors 
interpolate the experimental $e^-$--data linearly to the effective
energy of the $e^+$--data, $E_{e^+} = E_{e^-} - 2V_c$ with the
Coulomb potential $V_c$ and the relative normalization as
fit parameters. They find for $V_c$ a value close to the
surface and a normalization factor which is compatible with
the experimental uncertainties.
For the interpolation the large body of inclusive quasielastic 
data on $^{208}$Pb measured by Zghiche {\em{et al.}}
\cite{Zghiche} is used.
The authors conclude that a {\em modified} EMA with a
$V_c(r)=18.7$ MeV
close to the surface is a valid approach to correct cross
section data. 

However, this conclusion is highly questionable when the quality of
the data by Zghiche {\em{et al.}} is studied. The peculiar behavior
of the data is shown in Fig. 7, where the scaling function $F(y)$
(essentially the cross section but with the $q$-dependence of
the nucleon form factor and the kinematical broadening of the
quasielastic peak removed) is shown
(see \cite{Jourdan96a} for detailed definitions).
Although the electron energies progress in regular steps between
the various data sets, the scaling functions $F(y)$
display a curious behavior. The peak values
for $F(y)$ for two consecutive energies coincide and jump for the
following to a significantly lower value.
In Fig. 10 in \cite{Zghiche}, the staircase-like behavior of the
data is not visible, since cross sections are plotted, which depend
strongly on the initial electron energy $\epsilon_i^e$.
Additionally, the cross sections for 
$e_i^e=420$ MeV and $e_i^e=600$ MeV are missing there.
But the staircase-like
behavior has a detrimental effect upon the interpolation of the
electron data as shown by Gu\`eye {\em{et al.}}
Also, the $e^+$ and $e^-$ data of Gu\`eye {\em{et al.}} have been
$e^+$ beam emittance corrected and normalized to the $e^-$ data
shown in Fig. 7, hereby further affected by the staircase behavior. 
We therefore cannot consider
the experiment as a proof of the validity of the EMA.

We performed eikonal and
effective momentum calculations for
quasielastic electron (e) and positron (p) scattering on $^{208}$Pb for
a kinematics also used in the experiment of Gu\`eye
{\em{et al.}}, i.e. for
initial energy $\epsilon_i^{e}=\epsilon_i^{p}=420$ MeV and
scattering angle $\Theta_e=\Theta_p=60^o$. The calculations
for the positrons were
performed in a strictly analogous way as in the case of
electrons. Using an effective surface value of $17$ MeV 
for the repulsive (attractive) Coulomb potential
in the case of the positron (electron) EMA calculation,
it turns out that the eikonal approximation and the EMA
are clearly incompatible.
The observed discrepancy can be explained by
the assumption that the EMA underestimates the
defocusing of the positron waves and the focusing
of the electron wave in the nuclear volume.

For the sake of completeness, we also show in Fig. {\ref{fig10}}
experimental cross sections for positrons derived
from the total response function given in \cite{Gueye},
which are in satisfactory agreement with the theoretical
values shown in Fig. 9. 
As explained before, the difference between
experimental data and the theoretical cross sections
results from the simple nuclear model employed
(lack of high momenta in the corresponding spectral functions
and the absence of delta excitations and meson exchange current
contributions).
\begin{figure}
\begin{center}
        \includegraphics[width=9cm]{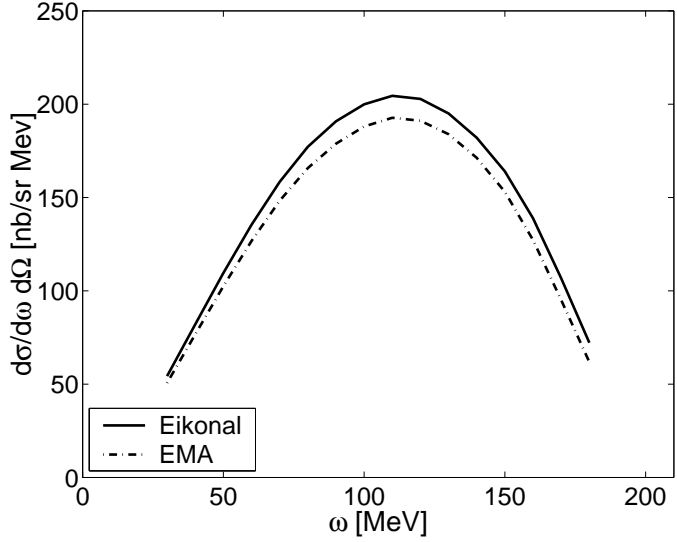}
        \caption{Theoretical cross sections for inclusive
         quasielastic electron scattering off $^{208}$Pb for
         initial electron energy $\epsilon_i^e=420$ MeV and
         scattering angle $\Theta_e=60^o$.}
        \label{fig8}
\end{center}
\end{figure}

\begin{figure}
\begin{center}
        \includegraphics[width=9cm]{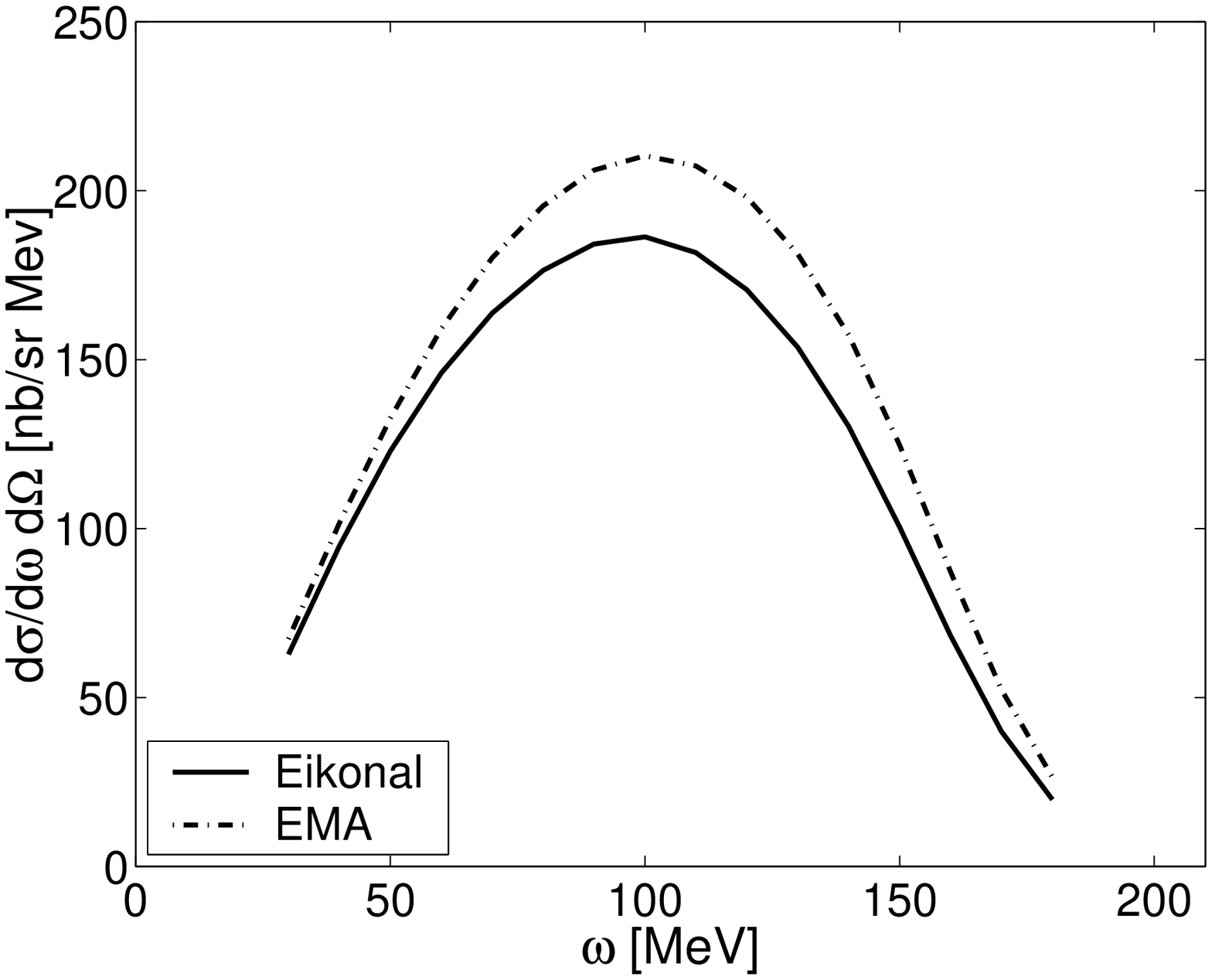}
        \caption{Theoretical cross sections for inclusive
         quasielastic positron scattering off $^{208}
         \mbox{Pb}$ for
         initial positron energy $\epsilon_i^p=420$ MeV and
         scattering angle $\Theta_p=60^o$.}
        \label{fig9}
\end{center}
\end{figure}

\begin{figure}
\begin{center}
        \includegraphics[width=9cm]{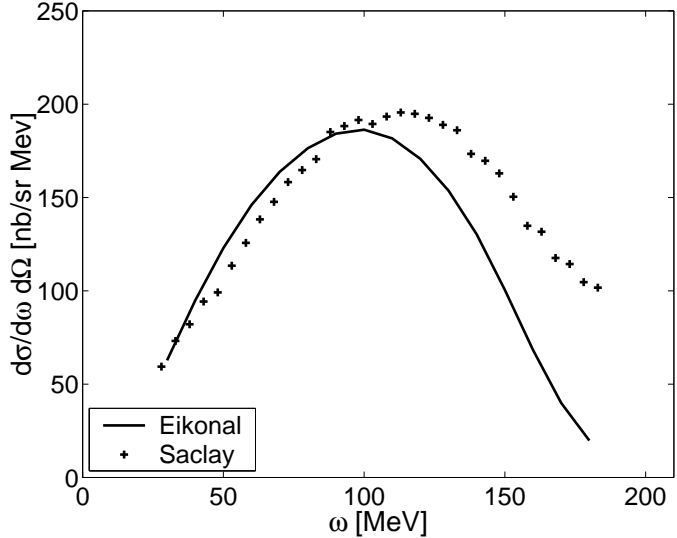}
        \caption{Experimental positron cross sections derived
         from the total response function given by Gu\`eye {\em{et
         al.}}, compared to the theoretical
         values shown in Fig. {\ref{fig9}} obtained from the single particle
         model.}
        \label{fig10}
\end{center}
\end{figure}

\section{Conclusions}
In the present paper we have investigated the role of the
Coulomb distortion in quasielastic electron nucleus scattering.
A reliable treatment of this distortion is needed in
particular for a determination of the longitudinal
response function and for an extrapolation of nuclear responses
to infinite nuclear matter.

We have developed an approximate description --the
eikonal approximation-- that is more transparent and
numerically easier to deal with than the exact treatments
(solution of the full Dirac equation)
previously employed \cite{Jin1,Udias,Udias2}.
At the same time, the eikonal approximation is much more
realistic than the effective momentum approximation often employed
in the absence of results from exact calculations.

We find that the eikonal results for the Coulomb
distortion are very close to the results of exact calculations.
We also find that the effective description
of the Coulomb distortion gives a rather
poor description for high values of the nuclear charge $Z$.

\section*{Acknowledgement}
We thank Gerhard Baur for many interesting and useful discussions
and Kyungsik Kim for a copy of the LEMA-code. This work was
supported by the Swiss National Science Foundation.

\section*{Appendix}
The potential energy of the electron according to Eq. (\ref{potential})
can be decomposed into three parts
$V(r)=V_1(r)+V_2(r)+V_3(r)$, where
\begin{displaymath}
V_1(r)=-\frac{\alpha Z}{(r^2+R^2)^{1/2}}, \quad
V_2(r)=-\frac{\alpha Z R^2}{2(r^2+R^2)^{3/2}},
\end{displaymath}
\begin{equation}
V_3(r)=-\frac{ 24 \alpha Z R^2 R' r^4}{25 \pi (r^2+R'^2)^4}.
\end{equation}
The regularized eikonal phase $\chi_1$ generated by $V_1$
is given above by Eq. (\ref{regularizedeikonal}).
Since $V_{2,3}$ decrease faster than $r^{-1}$ for large $r$,
their corresponding contributions to the total
eikonal phase for a particle incident parallel to the
z-axis with impact parameter $b$
\begin{equation}
\chi_{2,3}(\vec{r})=\chi_{2,3}(b,z)
=- \int \limits_{-\infty}^{z} dz' \, V_{2,3}(r'),
\quad r'^2=b^2+z'^2 \label{eikonalphases}
\end{equation}
need not be regularized.
The integrals in (\ref{eikonalphases}) are given by
\begin{equation}
\chi_2=\frac{\alpha Z}{2} \frac{R^2 (z+(r^2+R^2)^{1/2})}
{(b^2+R^2)(r^2+R^2)^{1/2}},
\end{equation}
\begin{displaymath}
\frac{25 \pi}{\alpha Z R^2 R'} \chi_3=\Biggl(  \frac{15}{2} \frac{b^4}{b'^7}
+\frac{3 b^2}{b'^5} + \frac{3}{2 b'^3} \Biggr) \arctan
\Biggl( \frac{z}{b'} \Biggr) - \frac{7 z}{(r^2+R'^2)^2}
\end{displaymath}
\begin{displaymath}
+\frac{4 b^4 z}{b'^2(r^2+R'^2)^3}+ \frac{5 b^4 z}
{b'^4(r^2+R'^2)^2} +\frac{15 b^4 z}{2 b'^6 (r^2+R'^2)}
-\frac{8 b^2 z}{(r^2+R'^2)^3}
\end{displaymath}
\begin{displaymath}
+\frac{2 b^2 z}{b'^2 (r^2+R'^2)^2}+ \frac{3 b^2 z}
{b'^4 (r^2+R'^2)} + \frac{4 b'^2 z}{(r^2+R'^2)^3}
+\frac{3z}{2 b'^2 (r^2+R'^2)}
\end{displaymath}
\begin{equation}
+\frac{3 \pi}{4} \frac{5 b^4 + 2b'^2 b^2 +b'^4}{b'^7},
\end{equation}
where $r^2=b^2+z^2$ and $b'=\sqrt{b^2+R'^2}$.
The coordinate independent definitions of
the parameters $z$, $b$, and $b'$ are $z=(\vec{k}_i \vec{r})/k_i$
and $b^2=r^2-z^2$, $b'=\sqrt{b^2+R'^2}$, where $\vec{k}_i$ is the
momentum of the incident electron.
The eikonal phase of the wave function of
the outgoing electron with momentum $\vec{k}_f$,
which must be added to the eikonal
phase of the incoming electron in the
matrix element of the calculated process,
can be obtained directly from the expressions above
by replacing $z=-(\vec{k}_f \vec{r})/k_f$ and using again
the definitions $b^2=r^2-z^2$, $b'=\sqrt{b^2+R'^2}$.

\end{document}